\documentclass[twocolumn,amsmath,amssymb,aps,prl,floatfix,nofootinbib,superscriptaddress]{revtex4}
\usepackage{graphicx,graphics,times,color}% Include figure files
\usepackage{bm}% bold math
\usepackage{longtable}
\def\be{\begin{equation}}
\def\ee{\end{equation}}
\def\ba{\begin{eqnarray}}
\def\ea{\end{eqnarray}}
\def\la{\langle}
\def\ra{\rangle}

\begin{document}
\title{Memory Effects in Spin-Chain Channels for Information Transmission}
\author{Abolfazl Bayat}
%\email{bayat@physics.sharif.edu}
\affiliation{Department of Physics and Astronomy, University College
London, Gower Street, London WC1E 6BT, United Kingdom}
\affiliation{Department of Physics, Sharif University of Technology,
P.O. Box 11365-9161, Tehran, Iran}
\author{Daniel Burgarth}
\affiliation{Computer Science Department, ETH Zurich, CH-8092 Zurich, Switzerland}
\author{Stefano Mancini}
\affiliation{Dipartimento di Fisica, Universita di Camerino, I-62032 Camerino, Italy}
\author{Sougato Bose}
\affiliation{Department of Physics and Astronomy, University College London, Gower St., London WC1E 6BT, UK}
\begin{abstract}
We investigate the multiple use of a ferromagnetic spin chain for
quantum and classical communications without resetting. We find that
the memory of the state transmitted during the first use makes the
spin chain a qualitatively different quantum channel during the
second transmission, for which we find the relevant Kraus operators.
We propose a parameter to quantify the amount of memory in the
channel and find that it influences the quality of the channel, as
reflected through fidelity and entanglement transmissible during the
second use. For certain evolution times, the memory allows the
channel to exceed the memoryless classical capacity (achieved by
separable inputs) and in some cases it can also enhance the quantum
capacity.
\end{abstract}
\date{\today}
 \maketitle Recently spin chains have been proposed as potential channels for short distance
quantum communications (See, for example, Refs.\cite{boseCP,danielthesis}).
The basic idea is to simply place the state to be transmitted at one
end of a spin chain initially in its ground state, allow it to propagate for a specific amount of time, and then receive it at the other end.
Generically, while propagating, the information will also inevitably disperse in the chain, and even when a transmission is considered complete
(i.e., the state is considered to have been received with some fidelity/probability),
some information of the state lingers in the channel. It
is thus assumed that a reset of the spin chain to its ground state is made after each transmission \cite{Giovannetti}. If, on the other hand, a
second transmission is performed through the channel without resetting, then the memory of the first transmission should affect the second
transmission. A spin chain channel without resetting is thus an
interesting physical model of a channel with memory \cite{Kretschmann}.\\
 In this paper, we show that a ferromagnetic spin chain used without
resetting is a very different channel than those studied so far in
the extensive literature of quantum channels with memory
\cite{macchiavello,Daems,macchiavello2,Arshed,memarzadeh,Datta,BDW+04,Hamada,Daffer,Bowen,Kretschmann,Bowen2,plenio,Arrigo}.
Firstly, the channels usually studied are those with the noise
during multiple uses being correlated with each other
\cite{macchiavello,Daems,macchiavello2,Arshed,memarzadeh,Datta,BDW+04},
but being independent of the transferred states. In our model,
however, the state transmitted during the first use modifies the
type of noise during the second use. Secondly, the noise is most
often assumed as Markovian correlated
\cite{macchiavello,Daems,macchiavello2,Arshed,memarzadeh,BDW+04},
while this is not the case for us. Thirdly, and most importantly,
the channel noises in our case stem from a physical model described
by a Hamiltonian. This should stimulate activity in calculating its
capacities. To this end, we also introduce a \textit{ memory
parameter} to quantify the amount of memory. This parameter depends
on the distance between the Kraus operators of the second use of the
channel with and without memory, so this method can be used to
quantify the amount of memory for those channels that admit a
description in terms of separate Kraus
 operators on different uses.\\
 There is also a very important practical issue which motivates our work. The
standard way of resetting the chain requires its interaction with a
zero temperature environment \cite{Giovannetti2} and this may open
up unnecessary avenues for decoherence. Thus one either resets
actively by performing a cooling sequence at the chain ends
\cite{danielthesis} or uses it several times without resetting which
automatically raises the question of the effect of memory of one
transmission on a subsequent transmission. Multiple usage of a chain
of two spins has been studied in \cite{rossini} to compute the rate
of information transmission, but
 using the swap operators on both spins, a chain of length $N=2$ removes the memory effects.
We will compare and contrast our results for the ferromagnetic
channel without resetting with some results that have emerged in the
recent literature
 \cite{macchiavello,Daems,macchiavello2,Arshed,memarzadeh,Datta,BDW+04,Hamada,Daffer,Bowen,Kretschmann,Bowen2,plenio,Arrigo}.\\
\begin{figure}
\centering
    \includegraphics[width=8cm,height=1.8cm,angle=0]{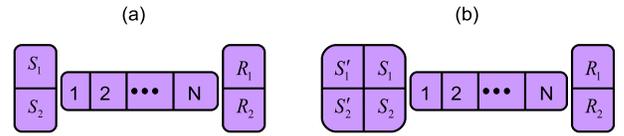}
    \caption{(Color online) Communication setup includes the sender and receiver registers in both side of the spin chain.
(a) Setup for state transfer and the classical capacity problem. (b)
Setup for entanglement distribution and the quantum capacity
problem.}
     \label{fig1}
\end{figure}
Let us consider a communication system like that of fig.
\ref{fig1}(a) which has a set of sender and receiver registers to
store the quantum input and output states respectively and a
ferromagnetic open spin chain as a quantum channel. The registers
are isolated from the channel, and the Hamiltonian $H_{ch}$ of the
chain commutes with $S_z$ (total spin in the $z$ direction) so the
number of the excitations in the channel is preserved through the
dynamics. Specifically we are going to consider a Heisenberg chain
with $N$ spins coupled by the Hamiltonian
$H_{ch}=-J\sum_{i=1}^{N-1}\sigma_i\cdot
\sigma_{i+1}-B\sum_{i=1}^{N}\sigma_i^z$ where $J$ and $B$ are the
coupling and magnetic field, respectively, and
$\sigma_i=(\sigma_i^x,\sigma_i^y,\sigma_i^z)$ is the vector of
 Pauli operators at site $i$.
 To transfer a quantum state from the register
 $S_k$ (we will restrict our attention to two uses of the channel, so $k=1,2$) to the register $R_k$ we put
the state in the channel by applying a swap operator $P_S(k)$ which
exchanges the state of the register $S_k$ and the first spin of the
channel $P_S(k)|\alpha\beta\ra_{S_k,1}=|\beta\alpha\ra_{S_k,1}$,
then we leave the spin chain to evolve for time $\tau_k$ and finally
the transmission is completed by applying another swap operator
$P_{R}(k)$ which exchanges the state of site $N$ with the register
$R_k$. The total operator to transfer the quantum state from the
sender register $S_k$ to the receiver register $R_k$ is
$W(k)=P_R(k)U(\tau_k)P_S(k)$. The initial state of the system is
$\rho(0)=\rho_{_{S}}(0) \otimes \rho_{ch}(0) \otimes\rho_R(0)$,
where $\rho_S(0)$ is an arbitrary initial density matrix of the
sender registers, $\rho_{ch}(0)=|\psi_{GS}\ra\la\psi_{GS}|$ is the
ground state of the chain and $\rho_R(0)$ contains all receiver
register spins in the states $|0\ra$. In the numerical analysis for
this paper we have used $N=4$. In fact, for $N=2$ the
 operations $P_S$ and $P_R$ exclude any memory effects \cite{rossini} while for $N=3$ the
quality of transmission is low \cite{boseCP}.\\
After the first transmission the total state is
$\rho(1)=W(1)\rho(0)W^\dagger(1)$ which is the case studied in
\cite{boseCP}. The received state in the register $R_1$ is
$\rho_{R_1}(1)=\sum_{i=1}^{N+1}M_i \rho_{_{S_1}}(0)M_i^\dag$, where
$M_i$'s are the following operators:
\begin{eqnarray}\label{krausampli}
    M_m=\left(%
\begin{array}{cc}
  0 & f_{m1}(\tau_1) \\
  0 & 0 \\
\end{array}%
\right),\  M_N=\left(%
\begin{array}{cc}
  1 & 0 \\
  0 & f_{N1}(\tau_1) \\
\end{array}%
\right),
\end{eqnarray}
with the index $m$ going from $1$ to $N-1$. In the above equation,
$f_{m1}(\tau_1)=\la \textbf{m}|U(\tau_1)|\textbf{1}\ra$, where
$|\textbf{m}\ra$ represents one flipped spin $|1\ra$ in site $m$ of
the channel and all the other spins in $|0\ra$. The operator
$M_{N+1}$ is a zero matrix which is included here for comparison
with the memory case later on. The effect of the operators $M_m$
($m=1,...,N-1$) can be combined into one operator, to show that the
chain acts as an amplitude damping channel \cite{boseCP}. Except the
case of perfect transfer, some information of the first state
remains in the state of the channel and the effect of channel is no
longer described by the Kraus operators (\ref{krausampli}). We
assume that in the first transmission the state of the sender
register $S_1$ is a general pure state
$r|0\ra+e^{i\phi}\sqrt{1-r^2}|1\ra$, but it is easy to generalize
the results to mixed input states. After the first transmission the
state of the channel can be calculated by tracing out the state of
the registers from $\rho(1)$. We obtain
\begin{equation}\label{rhoch2}
    \rho_{ch}(1)=p_0|\textbf{0}\ra \la
    \textbf{0}|+p_1|\psi_1\ra\la \psi_1|,
\end{equation}
where
\begin{eqnarray}\label{probabilities}
p_0&=&(1-r^2)|f_{N1}(\tau_1)|^2,\ \ \ \ \ p_1=1-p_0,\cr
|\psi_1\ra&=&\frac{1}{\sqrt{p_1}}(r|\textbf{0}\ra+\sqrt{1-r^2}e^{i\phi}\sum_{n=1}^{N-1}f_{n1}(\tau_1)|\textbf{n}\ra).
\end{eqnarray}
The state (\ref{rhoch2}) shows that with probability $p_0$ the
channel is in the state $|\textbf{0}\ra$ and acts like an amplitude
damping channel but there are some corrections with probability
$p_1$ due to the state $|\psi_1\ra$. To find the Kraus operators of
the channel with the state $|\psi_1\ra$ one can consider a general
density matrix in $S_2$ where the channel is in the state
$|\psi_1\ra$. By applying the operator $W(2)$ on the state of whole
system, the state of the register $S_2$ is transferred to the
register $R_2$ (albeit with a certain fidelity), so the Kraus
operators can be easily derived. We will write down the Kraus
operators in a certain way (for simplicity and interpretation),
though ours may not be the only way to write the Kraus operators for
the channel. Two of the Kraus operators of the channel with the
initial state $|\psi_1\ra$ are as in (\ref{krausampli}) multiplied
by the coefficient $\sqrt{\frac{1-r^2}{p_1}|f_{11}(\tau_1)|^2}$ and
the others are some matrices that we shall soon introduce. Thus we
can describe the effect of the channel with initial state
$|\psi_1\ra$ as a probabilistic effect, which means that with
probability $q=\frac{1-r^2}{p_1}|f_{11}(\tau_1)|^2$ the channel
affects the inputs like an amplitude damping channel with Kraus
operators (\ref{krausampli}) and with the probability $(1-q)$ the
effect of the channel is specified by the following Kraus operators:
\begin{eqnarray}\label{kraus-dec}
M'_{m}&=&\frac{1}{\sqrt{p_1-p_1q}}\left(%
\begin{array}{cc}
  A_m\sqrt{1-r^2}e^{i\phi} & f_{m1}(\tau_2)r \\
  0 & B_{mN}\sqrt{1-r^2}e^{i\phi} \\
\end{array}%
\right), \cr
M'_{N}&=&\frac{1}{\sqrt{p_1-p_1q}}\left(%
\begin{array}{cc}
  r & 0 \\
  A_N\sqrt{1-r^2}e^{i\phi} & rf_{N1}(\tau_2) \\
\end{array}%
\right),\cr
M'_{N+1}&=&\frac{1}{\sqrt{p_1-p_1q}}\left(%
\begin{array}{cc}
  0 & \sqrt{1-r^2}e^{i\phi}\sqrt{\sum_{k_1k_2}'|B_{k_1k_2}|^2} \\
  0 & 0 \\
\end{array}\right),\nonumber\\
\end{eqnarray}
where the index $m$ goes from 1 to $N-1$,
$\sum_{k_1k_2}'=\sum_{k_1=1}^{N-1}\sum_{k_2=k_1+1}^{N-1}$ and
$A_m=\sum_{n=2}^{N-1}f_{mn}(\tau_2)f_{n1}(\tau_1)$.
$B_{k_1k_2}=\sum_{n=2}^{N-1}f_{k_1k_2,Nn}(\tau_2)f_{n1}(\tau_1)$ is
the two excitation amplitude transition with $f_{pq,nm}=\la
\textbf{pq}|e^{-iHt}|\textbf{nm}\ra$, and $|\textbf{nm}\ra$ means
all the spins of the channel are in $|0\ra$ except the sites $n$ and
$m$. Notice that $B_{k_1k_2}$ includes physical interaction
(scattering) between the first and second state.\\
In order to get a complete description of the channel for the second
 use we know that with probability $p_0$ the state of the channel is $|\textbf{0}\ra$
(the spin chain is an amplitude damping channel) and with
probability $p_1q$ the state of the chain is $|\psi_1\ra$ but acts
as an amplitude damping channel. Thus with total probability
$p_0+p_1q$ the spin chain is an amplitude damping channel, otherwise
with the probability $p_1(1-q)$ the channel is in the state
$|\psi_1\ra$ and its effect is specified by the Kraus operators
(\ref{kraus-dec}). Therefore, we have
\begin{equation}\label{general_channel}
    \rho_{R_2}(2)=(p_0+p_1q)\xi_{AD}(\rho_{S_2}(0))+(p_1-p_1q)
    \xi_{Mem}(\rho_{S_2}(0)),
\end{equation}
where $\xi_{AD}$ is the amplitude damping evolution (\ref{krausampli}) and $\xi_{Mem}$ is the evolution
with Kraus operators (\ref{kraus-dec}).\\
If we consider the memory as a deviation of the channel effect from
the memoryless case, then to find a distance between the two
evolutions we can consider the distance between the Kraus operators
in the two cases. Thus, to quantify this deviation, the following
memory parameter is suggested:
\begin{equation}\label{memory parameter}
    \Delta=(p_1-p_1q) tr\{\sum_{m=1}^{N+1} (M'_m-M_m)^\dag (M'_m-M_m)\}.
\end{equation} Notice that we have multiplied
the summation of the distances in Eq. (\ref{memory parameter}) by
$p_1-p_1q$ which is the probability that this evolution takes place.
By substituting the exact form of the operators in \eqref{memory
parameter} for the case $\tau_1=\tau_2=\tau$, we arrive at
\begin{equation}\label{memory parameter2}
    \Delta/2 =(1-r^2)(1-|f_{11}|^2-|f_{N1}|^2)+(r-\sqrt{p_1-p_1q})^2.
\end{equation}\nonumber
It is clear that the memory parameter is dependent on the first
input of the chain as well as the channel parameter $\tau_1$. The
largest deviation from the memoryless case is given for $r=0$,
corresponding to the transmission of $|1\rangle$ on the first use.
In this case the maximum of $\Delta$ is
$4(1-|f_{N1}|^2-|f_{11}|^2)$. For $|f_{N1}(\tau_1)|=1$ we have
perfect transfer, and for $|f_{11}(\tau_1)|=1$ the first state is
swapped out by the sender into $S_2$.

\begin{figure}
\centering
    \includegraphics[width=8.5cm,height=4.5cm,angle=0]{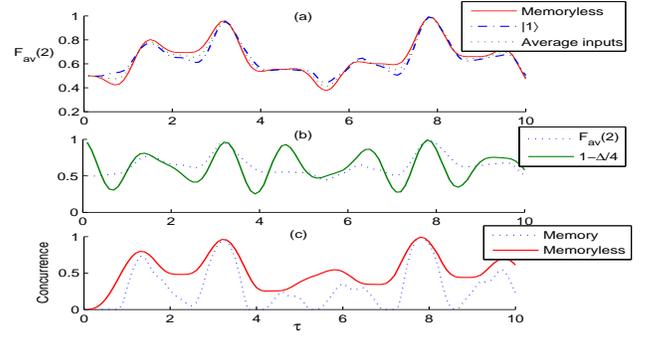}
    \caption{(Color online) As a function of evolution time $\tau$: (a) Average fidelity in the second use for memoryless
channel in comparison with the memory case when the state $|1\ra$ is
transferred and also with average pure input states in the first
use. (b) The average fidelity for the second use and the parameter
$1-\Delta/4$ after transferring the state $|1\ra$. (c) The
entanglement distribution for both the memory and memoryless
channel. }
     \label{fig2}
\end{figure}

 To compare the quality of
transmission we can compare the average fidelities. The average
fidelity in the $k$th use of the channel is $F_{av}(k)=\int
F(k)d\Omega$ where $F(k)= tr\{\rho_{S_k}(0)\rho_{R_k}(k)\}$ is the
fidelity of the $k$th transmission and the integration performed
over the surface of the Bloch sphere for all pure input states
$\rho_{S_k}(0)$. The total description of the channel in the second
use, Eq. (\ref{general_channel}) helps to compute the average
fidelity for the second transmission.  It is easy to show that,
\begin{eqnarray}\label{Fav}
F_{av}(2)&=&(p_0+p_1q)F_{av}(1)\cr
&+&\frac{1-r^2}{6}\sum_{m=1}^{N-1}2
Re(A_mB^*_{mN})-\frac{(1-r^2)|A_N|^2}{6}\cr
&+&\frac{2(1-r^2)}{3}(1-|f_{11}|^2-|f_{N1}|^2),
\end{eqnarray}
where
$F_{av}(1)=\frac{1}{2}+\frac{f_{N1}+f_{N1}^*}{6}+\frac{|f_{N1}|^2}{6}$
is the average fidelity for memoryless case, and we have used the
identity that
$\sum_{m=1}^{N}|A_m|^2=\sum_{k_1=1}^{N-1}\sum_{k_2=k_1+1}^{N}|B_{k_1k_2}|^2=1-|f_{N1}(\tau_1)|^2-|f_{11}(\tau_1)|^2$
to simplify the final result. In fig. \ref{fig2}(a) the average
fidelities for the second use of the channel has been plotted for
equal time evolutions $\tau_1=\tau_2=\tau$ (setting $J=1$). In this
figure the average fidelity for the memoryless case has been
compared with the case where the state $|1\ra$ has been transferred
in the first use and with the case of average inputs in the first
transmission. When the average fidelity of the first transmission
has a peak, which means almost perfect transmission, the next
transmission is also good. In non-optimal times when the first
transmission is not good the memory effect can improve the quality
of transmission. In fig. \ref{fig2}(b) the parameter $1-\Delta/4$
(we have used this parameter instead of $\Delta$ just for
simplicity) and the average fidelity for the second transmission
after sending the state $|1\ra$ in the first use, have been plotted
together. When $1-\Delta/4$ take its minimum it means that the
amount of the memory in the channel is high, so the average fidelity
in the second transmission has a low value because the state of the
channel is highly mixed and there is information from the previous
transmission in it. In the other case when the parameter
$1-\Delta/4$ has a peak it means that after the first transmission
the channel has been nearly reset to the initial ground state. But
in this case the average fidelity for the second transmission is not
necessarily high because the average fidelity also depends from the
time evolution $\tau$. For example in fig. \ref{fig2}(b) for
$\tau\simeq 4.6$ the memory has a low value but the average fidelity
is not high because of the non-optimal $\tau$. In this non-optimal
time, $|f_{11}|$ has a large value, which means that the information
is packed in the first spin and swapped out to the sender register,
so the chain reset to its ground state. The same happens for the
second transmission, so that the average fidelity is low.

\begin{figure}
\centering
    \includegraphics[width=7cm,height=3.0cm,angle=0]{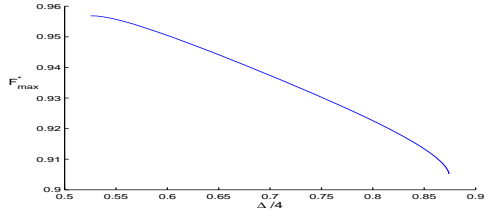}
    \caption{(Color online) Maximum of the average fidelity in the second use when $|1\ra$ is sent in the first use
    versus memory parameter $\Delta$.}
     \label{fig3}
\end{figure}

\begin{figure}
\centering
    \includegraphics[width=8cm,height=3.5cm,angle=0]{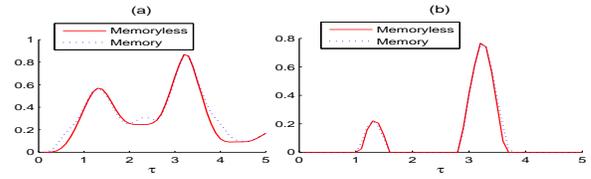}
    \caption{(Color online) (a) Holevo bound in memory spin chains in compare with classical capacity for separable input states.
(b) Coherent information for a special input states in memory channel in compare with quantum capacity for memoryless one.}
     \label{fig4}
\end{figure}

Another problem that can be compared for different uses of the
channel is the entanglement distribution. In this case the sender
registers are a set of pair registers like fig. \ref{fig1}(b). Dual
registers $S'_kS_k$ (k=1,2) contain a maximally entangled state. In
the first transmission the state $S_1$ is transferred to the
register $R_1$ to create an entangled pair (not necessarily maximal)
between $S'_1R_1$. In the second transmission, without resetting the
chain, the state of $S_2$ is transferred to the register $R_2$ to
create the entanglement between $S'_2R_2$. In fig. \ref{fig2}(c) the
concurrence as a measure of entanglement \cite{Wootters} for the
states $\rho_{S'_1R_1}$ (memoryless) and $\rho_{s'_2R_2}$ (memory
case) has been plotted. It shows that the effect of memory is always
destructive. The peaks of
entanglement are located at times where nearly perfect transmission happens.\\
Let us now discuss the dependence of the fidelity on $\Delta$. As
shown above the quality of state transmission in the second use of
the channel depends on the time evolution $\tau_1$ as well. We chose
a range of $3.3\leq\tau_1\leq3.9$ such that the memory parameter is
increasing for the case that the state $|1\ra$ is transferred in the
first use. For each value of $\tau_1$ we have compared the maximum
average fidelity in a long range of $\tau_2$. In fig. \ref{fig3} we
have plotted this maximum value of the average fidelity $F_{av}^*$
in the second transmission versus $\Delta$. Figure \ref{fig3} is
very interesting because it shows that the average fidelity is
decreasing when $\Delta$ is increased. This shows that the remaining
probability amplitude in the chain has a destructive effect on the
quality
of transfer in the second use of the chain.\\
Finally we investigate whether the memory effect (taking equal
evolution times $\tau_1=\tau_2=\tau$ for simplicity) can enhance
either the quantum capacity or the single-shot classical capacity
which are both known for the memoryless (amplitude damping) channel
\cite{Giovannetti}. As we will show below, such enhancement is
indeed possible, and can be demonstrated even without explicitly
calculating the capacities. We compare the Holevo bound for a
special equiprobable bipartite input states in memory channel with
the classical capacity of separable input states in memoryless
channels \cite{Giovannetti}. Assume that all the four possible
equiprobable classical input data are encoded into a special kind of
input states,
\begin{eqnarray}\label{inputstates}
    |\phi_1(\theta)\ra &=&\cos\theta|++\ra+\sin\theta|--\ra\cr
    |\phi_2(\theta)\ra &=&\sin\theta|++\ra-\cos\theta|--\ra\cr
    |\phi_3(\theta)\ra &=&\cos\theta|+-\ra+\sin\theta|-+\ra\cr
    |\phi_4(\theta)\ra &=&\sin\theta|+-\ra-\cos\theta|-+\ra,
\end{eqnarray}
where $|\pm\ra=(|0\ra\pm|1\ra)/\sqrt{2}$ and all these sates vary
from separable states ($\theta$=0) to the maximally entangled one
($\theta=\pi/4$). In fig. \ref{fig1}(a) one can prepare any of
states $|\phi_i\ra$ in registers $S_1$ and $S_2$ and by applying the
operator $W(2)W(1)$ this state is received as the state $\varrho_i$
in registers $R_1$ and $R_2$. The Holevo bound for input states
(\ref{inputstates}) per use is
$C(\tau,\theta)=\frac{1}{2}\{S(\sum^4_{i=1}p_i\varrho_i)-\sum_{i=1}^4p_iS(\varrho_i)\}$,
where $p_i=1/4$ and $S(\varrho)=-tr{\varrho \ln\varrho}$ is the Von
Neumann entropy of the state $\varrho$. To find the optimal input
states one can maximize $C(\tau,\theta)$ over the parameter
$\theta$. Surprisingly, the maximum
$C_{max}(\tau)=max_{\theta}\,C(\tau,\theta)$ is always achieved by
separable states ($\theta=0$). In fig. \ref{fig4}(a) we have plotted
the $C_{max}(\tau)$ and also the real capacity of memoryless channel
with separable input states \cite{Giovannetti} in terms of $\tau$.
The memory helps to increase the classical capacity in non-optimal
times. These results for spin chains
are analogous to those of memory dephasing channel \cite{Arrigo}.\\
The coherent information as a lower bound for quantum capacity is
$I=S(\xi(\rho))-S(I\otimes\xi(|\phi\ra\la \phi|))$, where $\rho$ is
the input and$|\phi\ra$ is a purification of $\rho$. In fig.
\ref{fig1}(b) consider two maximally entangled states in registers
$S'_1S_1$ and $S'_2S_2$ so the states of unprimed sender registers
are $\rho_{S_1,S_2}=I/2\otimes I/2$. These two states are
transferred through the chain by $W(2)W(1)$ and we can consider two
maximally entangled states in registers $S'_1S_1$ and $S'_2S_2$ as a
purification of transferred states. In fig. \ref{fig4}(b) we compare
the quantum capacity of \cite{Giovannetti} with the coherent
information per use in our model. We see that though the effect is
small, there are certain memory channels ( i.e., certain $\tau$) for
which even a lower bound to the true quantum capacity exceeds the
memoryless quantum capacity.

In conclusion, we have given a characterization of the behavior of a
spin chain without resetting. It provides an interesting example of
a quantum memory channel, where the memory of the state transmitted
during the first use produces a qualitatively different channel in
the second use.

 We have found the relevant
Kraus operators for this model and we have introduced a parameter to
quantify the amount of memory in the channel which has broader
applicability even outside the domain of spin chain channels.

We have shown that the memory effect can enable one to exceed the
known classical capacity for separable inputs and the quantum
capacity of the memoryless channel. Our study might pave the way for
the computation of the full capacities of such a spin chain channel
with memory.

\smallskip
S.B. is supported by the EPSRC, through which a part of the stay of
A.B. at UCL is funded, the QIP IRC (GR/S82176/01), the Wolfson
Foundation, and the Royal Society of Science and Technology. A.B.
thanks the British Council in Iran for financial support. D.B. and
S.M. are grateful to S.B. for the hospitality at UCL.

\end{document}